\documentclass[aps,prd, floatfix,twocolumn,preprintnumbers,superscriptaddress,showpacs]{revtex4-1}
\usepackage{graphicx}
\usepackage{slashed,color}
\usepackage{amsmath,amsfonts,bm}
\usepackage{amssymb}
\usepackage[T1]{fontenc}

\newcommand{\nn}{\nonumber \\ }
\newcommand{\varv}{v}

\begin{document}

%\begin{frontmatter}

\preprint{JLAB-THY-19-3039}

\author{A.~V.~Radyushkin}
\address{Old Dominion University, Norfolk,
             VA 23529, USA}
\address{Thomas Jefferson National Accelerator Facility,
              Newport News, VA 23606, USA
}

\title{Generalized Parton Distributions and Pseudo-Distributions}

\begin{abstract}

%\bigskip

We derive one-loop matching relations for the Ioffe-time 
distributions related to the pion distribution amplitude (DA)
and generalized parton distributions (GPDs).
They  are obtained from a universal 
expression for the one-loop correction  in an operator form, and  
 will be used in the ongoing lattice calculations 
of the pion DA and GPDs  based on the parton
pseudo-distributions approach.

%\vspace{5mm} 

%Keywords: Parton distribution  functions; Transverse  momentum; Quasi-distributions
              
\end{abstract}

%\pacs{11.10.-z,12.38.-t,13.60.Fz}
\maketitle

%\end{frontmatter}

\section{
 Introduction}

Extraction of  parton distribution functions (PDFs) 
from lattice simulation attracts now a considerable interest and efforts
(for a recent review see Ref.  \cite{Cichy:2018mum}).
An intensive development in this field has started 
with the paper by X. Ji  \cite{Ji:2013dva},
who proposed the concept of parton quasi-distributions 
(quasi-PDFs)  formalized later 
within  a general  framework of the Large Momentum Effective Theory (LaMET) 
\cite{Ji:2014gla}.  The basic idea of Ref.  \cite{Ji:2013dva}
(preceded in  Refs. \cite{Detmold:2005gg,Braun:2007wv})
to study equal-time correlators is also used in the ``good lattice cross sections''
approach  \cite{Ma:2014jla,Ma:2017pxb}
and in the pseudo-PDF approach \cite{Radyushkin:2017cyf,Radyushkin:2017sfi,Orginos:2017kos}.

The conversion of the Euclidean-space lattice data into
the standard  light-cone PDFs is performed with the help
of the  {\it matching relations}.
In the quasi-PDF approach, such relations were derived 
for the usual PDFs \mbox{\cite{Ji:2013dva,Xiong:2013bka,Ji:2015jwa,Izubuchi:2018srq,Wang:2019tgg},}
the pion distribution amplitude  (DA) \cite{Ji:2015qla} and generalized parton distributions (GPDs) 
\cite{Ji:2015qla,Xiong:2015nua,Liu:2019urm}. 
The matching relations between the pseudo-PDFs and the usual light-cone 
PDFs were discussed in Refs. \cite{Ji:2017rah,Radyushkin:2017lvu,Radyushkin:2018cvn,Zhang:2018ggy,Izubuchi:2018srq}.

It should be noted that, in all the papers mentioned above,    the derivation of the matching relations
was based on   separate 
calculations  of the relevant one-loop Feynman diagrams.
However, as pointed out in our  paper \cite{Radyushkin:2017lvu},
the one-loop correction in the coordinate-representation approach of 
Ref. \cite{Balitsky:1987bk}
 may be calculated in the operator form, 
 i.e., without specifying the   matrix element characteristic of   a particular 
 parton distribution. 
 
The  diagram by diagram results of such  one-loop calculation 
for a nonsinglet quark operator 
 are given 
 in Ref. \cite{Radyushkin:2017lvu}, and they were 
 used there to obtain the matching  relations between the non-singlet 
 pseudo-PDF and the corresponding PDF. 
 It was also stated there that the same  result obtained on the operator level 
 may be used to derive matching relations for the pion distribution amplitude
 and the non-singlet  generalized parton distributions.

 It is the goal of the present paper to describe the derivation
 of these matching relations. They  can be used in
 future lattice extractions of the pion DA and non-singlet GPDs 
 within the pseudo-PDF approach.

To make the paper self-contained, we describe in   \mbox{Sec. II}   the derivation
of the known matching relations for non-singlet PDFs.
 In Sec,  III, we  derive matching relations for the pion distribution amplitude.
 The matching relations for non-singlet GPDs are  derived in Sec. IV. 
 Sec. V    contains the  summary of the paper.

 %\setcounter{equation}{0}   {\it Parton distributions} 

%{\it DIS }

%\newpage

 \setcounter{equation}{0}  

\section{Matching conditions in the coordinate space}

\subsection{Operators and parton distributions} 

 In the present paper,  we will consider  the  valence parton distribution functions,
 the pion distribution amplitude (DA) and non-singlet  generalized parton distributions.
 They all are 
 given by  matrix elements  of  non-singlet operators of a generic form  
    \begin{align}
 { \cal O}^\alpha  (z) \equiv   \bar \psi (0) \,
 \Gamma^\alpha \,  { \hat E} (0,z; A) \psi (z) \  , 
\end{align}
where $\Gamma^\alpha = \gamma^\alpha$ or $  \gamma^\alpha \gamma_5$.
The factor 
${ \hat E}(0,z; A)$ is  the  standard  $0\to z$ straight-line gauge link 
 in the quark (fundamental) 
 representation
 \begin{align}
{ \hat E}(0,z; A) \equiv P \exp{ \left [ ig \,  z_\nu\, \int_0^1dt \,  \hat  A^\nu (t z) 
 \right ] }    \  . 
 \label{straightE}
\end{align}

In particular, studying the parton distribution functions, we deal with the  forward 
matrix elements 
 \begin{align}
M^\alpha (z,p) \equiv  \langle  p |  {\cal O}^\alpha (z)  | p \rangle 
\label{pdf}
 \end{align}
between the  hadronic states $ | p \rangle $ with momentum $p$.
 By Lorentz invariance, $M^\alpha (z,p)$ may be represented as a sum of two structures  
     \begin{align}
{ M}^\alpha  (z,p) =2 p^\alpha  {\cal M} (-(zp), -z^2) + z^\alpha  {\cal M}_z (-(zp),-z^2)
\ 
\end{align}
involving the amplitudes depending on two Lorentz scalars:
the interval  $z^2$ and  the invariant $(pz) \equiv -\nu$, 
the Ioffe time \cite{Ioffe:1969kf}. 

The twist-2 PDF is determined by the {\it Ioffe-time pseudo-distribution} 
$ {\cal M} (\nu, -z^2)$,
while ${\cal M}_z (\nu,-z^2)$ is a purely higher-twist 
contamination. It can be eliminated by an appropriate choice
of $z$ and $p$. The usual  way to define twist-2 PDF is to 
use $z$ in a purely ``minus'' light-cone direction, i.e., $z=z_-$ and 
take $\alpha =+$. To exclude ${\cal M}_z $ in    lattice calculations, one 
may use $z=z_3$ and $\alpha=0$,  as suggested in Ref. \cite{Radyushkin:2016hsy}.
We will follow this prescription for all  the parton distributions that 
we consider in  the present paper.

\subsection{One-loop correction in the operator form}  

The one-loop correction to $ { \cal O}^0  (z_3)$ was  calculated 
in the operator form in Ref. \cite{Radyushkin:2017lvu}, and  
is given by 
 \begin{align} 
\delta & {\cal O}^0 (z_3)   =  -  \frac{\alpha_s}{2\pi} \, C_F
\,\int_0^1  du \int_0^{1-u}  d\varv \, \bar \psi (uz_3) \Gamma^0 \psi (\bar \varv z_3) 
  \nn & \times  \left \{  \left ( \delta (\varv) \left[ \frac{\bar u }{u }\right ] _+
 + \delta (u) \left[ \frac{\bar \varv }{\varv }\right ] _+  +1  \right ) 
%  \times 
\ln \left [ z_3^2\mu_{\rm IR}^2 \frac{ e^{2\gamma_E+1}}{4}  \right ] \right. 
 \nn   & \left. \hspace{1cm} 
+    2   \left (\delta (\varv) \left[ \frac{\ln u }{u }\right ] _+
 + \delta (u) \left[ \frac{\ln\varv }{\varv }\right ] _+  -1 \right )
 \right. \nn   & \hspace{3cm} 
 +Z (z_3)  \delta (u) \delta (\varv) 
\Bigr \}  \  .
\label{Oha}
 \end{align}
 Here we use the notation $\bar \varv =1-\varv, \bar u =1-u$, etc. 
    The plus-prescription at zero  is defined by
   \begin{align} 
 \,\int_0^1  du \left[ \frac{\bar u }{u }\right ] _+  F(u)  =    \,\int_0^1  du \, \frac{\bar u }{u } \,  [F(u)-F[0] \ ,
    \label{Plus}
 \end{align}
 assuming that $F(0)$ is finite.
 
 In our result (\ref{Oha}),  we have used the dimensional regularization for 
 collinear singularities, and applied the $\overline{\rm MS}$ scheme 
 subtraction with  $\mu_{\rm IR}$ serving as the scale  parameter.  
 
 The function $Z (z)$ accumulates information about local corrections 
 associated  with the ultraviolet-divergent contributions. This function is also known
 (see  Ref. \cite{Izubuchi:2018srq}),
 but, in the pseudo-PDF approach,  we do not need   its explicit form.  As we will see,
     such terms cancel 
 when one forms the reduced Ioffe-time pseudo-distributions.
 
 In Feynman gauge, the terms containing $\delta (u)$ or $\delta (v)$ 
 in the integrand of Eq.(\ref{Oha}) are produced by vertex diagrams,
 while the $u,v$-independent terms come from the box diagram
 (see Ref. \cite{Radyushkin:2017lvu}).
 So, we will use sometimes ``vertex'' and ``box'' to refer to these two types of contributions. 
 
\subsection{Matching for parton distribution functions}

In the PDF case, the one-loop correction to $ { M}^0 (z_3,p)$  is given by 
the forward matrix element 
$\langle  p |\delta {\cal O}^0 (z_3)  | p \rangle $.  
Using translation invariance, the ``vertex'' terms containing $\delta(u)$ or $\delta (v)$ 
are trivially reduced  to one-dimensional integrals involving, say,  
$(\bar u/u)_+ {\cal M}_0( \bar u \nu)$ or $(\bar v/v)_+ {\cal M}_0( \bar v \nu)$.
Changing $u$ or $v$ to a common variable $1-w$,
we get the \mbox{$w$-integral} of $2(w/\bar w)_+ {\cal M}_0( w  \nu)$ 
with the plus-prescription at $w=1$.
For the ``box'' terms having a $u,v$-independent integrand, we get
   \begin{align} 
 \,  &
 \,\int_0^1  du \int_0^{1-u}   d\varv \, 
 {\cal M}_0 ((1-u-\varv) \nu)  \nn & =\int_0^1 dw\, (1-w) {\cal M}_0 (w \nu)
     \ .
 \end{align} 
We can represent $(1-w)$ 
 as the sum 
  of the term $(1-w)_+$ 
 that has the  plus-prescription 
 at $w=1$ and  the delta-function term  $ \frac12 \delta(\bar w)$ that we add to  $Z(z_3)$, 
denoting   the  changed $Z$-function  by   $\widetilde Z(z_3)$.
As a result, we have 
 \begin{align} 
{\cal M}(\nu,z_3^2) &  = \Biggl [1-  \frac{\alpha_s}{2\pi} \, C_F    \widetilde Z (z_3) \Biggr ]   {\cal M}_0(\nu)
\nn & 
-\frac{\alpha_s}{2\pi} \, C_F
\,\int_0^1  dw 
\Biggl  \{
   \frac{1+w^2 }{1- w }
\ln \left ( z_3^2\mu_{\rm IR}^2\frac{ e^{2\gamma_E+1}}{4}  \right ) 
 \nn   & 
+    4    \frac{\ln (1-w) }{1-w }
   -2(1-w)  \Biggr \}_+  {\cal M}_0( w \nu) 
 \  .
\label{deltaM}
 \end{align}

 The structure of Eq. (\ref{Oha})  implies a  scenario in which  the $z_3^2$-dependence
 at   short distances 
 is determined by the ``hard'' logarithms
 $\ln z_3^2$ generated from the  initially ``soft'' distribution ${\cal M}_0(\nu,z_3^2)$
 having  only a polynomial dependence on $z_3^2$ that is negligible
 for small $z_3^2$. For this reason, we skip 
 the \mbox{$z_3^2$-dependence}  in the argument of ${\cal M}_0$-functions,
 leaving just their $\nu$-dependence.
  The combination  
  \begin{align} 
B(w)  =&
       \left [\frac{1+w^2} {1-w}   \right ]_+
  \ 
     \label{V1}
  \end{align}   
  is   the non-singlet 
  Altarelli-Parisi (AP)  evolution kernel \cite{Altarelli:1977zs}. 
  The latter is usually defined  for PDFs, i.e.,  in the momentum-fraction space. However, 
 introducing the pseudo-PDF ${\cal P}(x,z_3^2)$  \cite{Radyushkin:2017cyf}  by 
\begin{align} 
{\cal P}(x,z_3^2)   =\frac{1}{2 \pi}  \int_{-\infty}^{\infty}  d\nu \, e^{-i x \nu}  \, {\cal M}(\nu , z_3^2)     
  \  ,
\label{fxMnu}
\end{align}  
or by the  inverse transformation 
  \begin{align}
  {\cal M} (\nu,z_3^2) =
   \int_{-1}^1 dy \,  \, 
e^{iy\nu} \, {\cal P}(y,z_3^2)  \   , 
 \label{twist2par0}
\end{align}
with ${\cal P}(x,0)\equiv f(x)$ being the  usual PDF, 
we see that the $\ln z_3^2$ part of Eq. (\ref{deltaM}) 
converts into
 \begin{align} 
&{\cal P}(x,z_3^2)  = f(x) \nn &  
-\frac{\alpha_s}{2\pi} \, C_F\, \ln \left ( z_3^2 \right )  
\,\int_0^1  dw  B(w)\int_{-1}^1dy \,  \delta (x-w y) f(y) + \ldots  \nn
&= 
 f(x)
-\frac{\alpha_s}{2\pi} \, C_F\, \ln \left ( z_3^2 \right )  
\int_{-1}^1\frac{dy}{|y|}  \, B(x/y)   f(y) +\ldots 
 \  ,
\label{PAP}
 \end{align}
which has the standard form of the DGLAP (for Dokshitzer-Gribov-Lipatov-Altarelli-Parisi 
\cite{Gribov:1972ri,Altarelli:1977zs,Dokshitzer:1977sg})   evolution equation. 

The next step is to introduce 
the reduced Ioffe-time pseudo-distribution  
 \begin{align}
{\mathfrak M} (\nu, z_3^2) \equiv \frac{ {\cal M} (\nu, z_3^2)}{{\cal M} (0, z_3^2)} \  
 \label{redm0}
\end{align}
of Refs. \cite{Radyushkin:2017cyf,Radyushkin:2017sfi,Orginos:2017kos}. 
When the momentum $p$ is also oriented  in the $z_3$ direction,
i.e., $p=\{E, 0_\perp,p_3\}$, the function ${\cal M} (0, z_3^2)$ corresponds to the  ``rest-frame'' $p_3=0$ distribution.
 According to Eq. (\ref{deltaM}), it is given by
  \begin{align} 
{\cal M}(0,z_3^2) &  =  {\cal M}_0(0)    \Biggl  [  1-  \frac{\alpha_s}{2\pi} \, C_F \widetilde Z (z_3) 
  \Biggr ]
 \  .
\label{deltaM0}
 \end{align}
 As a result, the $\widetilde Z(z_3)$ terms disappear from    the ${\cal O}(\alpha_s)$  correction to the ratio
 $ {\cal M} (\nu, z_3^2)/{\cal M} (0, z_3^2)$, and 
we have 
  \begin{align} 
{\mathfrak M}(\nu,z_3^2)  & =  {\mathfrak M}_0(\nu) 
 \nn   &- 
\frac{\alpha_s}{2\pi} \, C_F 
\,\int_0^1  dw   
\Biggl  [
 \frac{1+w^2 }{1- w }\, 
\ln \left ( z_3^2 \mu_{\rm IR}^2\frac{ e^{2\gamma_E+1}}{4}  \right ) 
 \nn   &% \hspace{1.3cm}
+    4    \frac{\ln (1-w) }{1-w }
   -2(1-w)  \Biggr ]_+{\mathfrak  M}_0( w \nu) 
 \  .
\label{ITDm}
 \end{align}
 Such a cancellation of ultraviolet terms for ${\mathfrak M}(\nu,z_3^2) $ will persist in higher $\alpha_s$ orders,
 reflecting the multiplicative renormalizability of the ultraviolet divergences of ${\cal M}(\nu,z_3^2)$ \cite{Ji:2017oey,Ishikawa:2017faj,Green:2017xeu}.

 A similar calculation can be performed for the light-cone 
 {\it Ioffe-time distribution}   ${\cal I} (\nu, \mu^2)$   \cite{Braun:1994jq}  obtained by taking $z^2=0$ 
 in ${\mathfrak M}(\nu,-z^2)$ and regularizing the resulting light-cone singularities 
using dimensional regularization and the $\overline{\rm MS}$ subtraction
specified by a factorization scale $\mu$. The result is 
  \begin{align} 
{\cal  I}(\nu,\mu^2)   =  {\mathfrak M}_0(\nu) 
 %\nn   &
 - & 
\frac{\alpha_s}{2\pi} \, C_F 
\,\int_0^1  dw   
\Biggl  [
 \frac{1+w^2 }{1- w }  \Biggr ]_+ \, 
 \nn & \times 
\ln \left ( \mu_{\rm IR}^2/\mu^2 \right ) 
{\mathfrak  M}_0( w \nu) 
 \  .
\label{LCITD}
 \end{align}
 Combining Eqs. (\ref{ITDm}) and  (\ref{LCITD})
 gives the matching condition for the light-cone ITD
  \cite{Ji:2017rah,Radyushkin:2017lvu,Radyushkin:2018cvn,Zhang:2018ggy,Izubuchi:2018srq} 
   \begin{align} 
   {\cal  I}(\nu,& \mu^2) = 
{\mathfrak M}(\nu,z_3^2)  
+
\frac{\alpha_s}{2\pi} \, C_F 
\,\int_0^1  dw   \, {\mathfrak  M}( w \nu,z_3^2)  
\Biggl  [
 \frac{1+w^2 }{1- w }\, 
 \nn & \times
\ln \left ( z_3^2 \mu^2\frac{ e^{2\gamma_E+1}}{4}  \right ) 
 %\nn   &% \hspace{1.3cm}
+    4    \frac{\ln (1-w) }{1-w }
   -2(1-w)  \Biggr ]_+
 \  
\label{Mtch}
 \end{align}
that allows to get $  {\cal  I}(\nu, \mu^2) $ from lattice data  on 
 ${\mathfrak  M}(  \nu,z_3^2) $.
By definition   \cite{Braun:1994jq},  the light cone ITD $  {\cal  I}(\nu, \mu^2) $  is related to the  PDF  ${f}(x,\mu^2)$
by 
   \begin{align}
  {\cal I} (\nu,\mu^2) =
   \int_{-1}^1 dx \,  \, 
e^{ix\nu} \, {f}(x,\mu^2)  \   .  
 \label{If}
\end{align}
 Thus, ${f}(x,\mu^2) $ is formally given by the inverse transformation 
 \begin{align} 
{f}(x,\mu^2)   =\frac{1}{2 \pi}  \int_{-\infty}^{\infty}  d\nu \, e^{-i x \nu}  \, {\cal I}(\nu , \mu^2)     
  \  .
\label{fI}
\end{align}  

However, lattice calculations  provide ${\cal I}(\nu , \mu^2)$ in a rather limited range 
of $\nu$,  which   makes taking  this Fourier transform rather tricky (see Ref. \cite{Karpie:2019eiq} 
for a detailed discussion). 
  An easier way was proposed in our paper \cite{Radyushkin:2017cyf}. The  idea 
  is to assume some parametrization for ${f}(x,\mu^2) $  similar to those  used 
  in global fits (see, e.g., Ref. \cite{Accardi:2016qay}), and to  incorporate   
 Eq. (\ref{If}) to fit its parameters using the lattice data for $ {\cal I}(\nu , \mu^2)   $.
 
 An equivalent realization of this idea (similar to that of Ref. \cite{Cichy:2019ebf})
 is
 to represent   ${\mathfrak M}(\nu,z_3^2)  $ in terms of $ {\cal  I}(\nu, \mu^2)$
 (see, e.g., Eq. (5.1) in Ref. \cite{Radyushkin:2018nbf}),  
 which, in turn is  written through its definition (\ref{If}) as a Fourier transform of PDF
    \begin{align} 
{\mathfrak M}(\nu,z_3^2)    &  = 
  \int_{-1}^1 dx \,  \, 
e^{ix\nu} \, {f}(x,\mu^2) 
-
\frac{\alpha_s}{2\pi} \, C_F 
\ \int_{-1}^1 dx \, {f}(x,\mu^2)     \nn & \times  \, \int_0^1  dw   \,
e^{ixw \nu} \,
\Biggl  [
 \frac{1+w^2 }{1- w }\, 
\ln \left ( z_3^2 \mu^2\frac{ e^{2\gamma_E+1}}{4}  \right ) 
 %\nn   &% \hspace{1.3cm}
  \nn & 
+    4    \frac{\ln (1-w) }{1-w }
   -2(1-w)  \Biggr ]_+
   \nn   & 
   \equiv  \int_{-1}^1 dx \,  \left [e^{ix\nu}  - \frac{\alpha_s}{2\pi} \, C_F 
   R( x \nu, z_3^2 \mu^2) \right] \,  {f}(x,\mu^2)   \ .
 \  
\label{MtchI}
 \end{align}
 The kernel $R(x \nu, z_3^2 \mu^2 )$, introduced in the equation above, 
  may be calculated as a  closed-form expression and  is given by
  \begin{align} 
& R(y, z_3^2 \mu^2) =  \frac1{ y^2} \Biggl  \{e^{iy} \Bigl [2y ^2
 \{\text{Ci}(y)-i
   \text{Si}(y)-\ln (y)\}
     \nn &  +y [(3-4 \gamma_E ) y 
   +4 i]/2-1 \Bigr ] 
 - i y +1 \Biggr \} \ln \left ( z_3^2 \mu^2\frac{ e^{2\gamma_E+1}}{4}  \right ) 
 \nn &\hspace{2cm} + 4 i y e^{i y} \,
   _3F_3(1,1,1;2,2,2;-i y) \nn & \hspace{2cm}   -\frac2{y^2} \Bigl [1+ i y-e^{i y} \left(1+y^2/2
  \right) \Bigr ] \, ,
\label{MtchR}
 \end{align}
 where $\text{Ci}(y)$ and $\text{Si}(y)$  are the  integral cosine and sine functions, and $_3F_3(1,1,1;2,2,2;-i y)$
 is a hypergeometric function. 
 
 One may split $f(x)$ in its symmetric $f^+(x)$ and antisymmetric  $f^-(x)$ parts.   
 For positive $x$,  they are  related to the quark $f_q (x)$ and antiquark $f_{\bar q} (x)$ distributions
 by $f^+(x)=f_q (x) - f_{\bar  q} (x)$ and $f^-(x)=f_q (x) + f_{\bar q } (x)$, respectively 
 (see, e.g., Ref. \cite{Orginos:2017kos}).
 Then the  real part of $ R(y, z_3^2 \mu^2)$ generates the real part of 
 ${\mathfrak M}(\nu,z_3^2) $ from $f^+(x)$, while 
 the  imaginary part of $ R(y, z_3^2 \mu^2)$ connects  the  imaginary  part of 
 ${\mathfrak M}(\nu,z_3^2) $ with  $f^-(x)$. 
 
 Thus, assuming some parameterizations for the $f^{\pm}(x,\mu^2)$ distributions,
 one can  fit  their  parameters and $\alpha_s$ using  Eqs. (\ref{MtchI}), (\ref{MtchR})  and  the  lattice data for ${\mathfrak M}(\nu,z_3^2)$.

 \section{Matching for pion distribution amplitude}

 \subsection{Definition and general properties}
 
    The  pion  distribution amplitude,  initially   introduced in our 1977 
    paper (see  Ref. \cite{Radyushkin:1977gp} 
    and also Ref. \cite{Lepage:1980fj}, where a similar object was introduced within 
    the  light-front quantization formalism), may be defined  using the matrix element  
 \begin{align}
M^\alpha (z,p)=  \langle  0 | \bar \psi (0) \,
 \gamma^\alpha \, \gamma_5 { \hat E} (0,z; A) \psi (z)    | p \rangle \ , 
\label{DA}
 \end{align}
where  $ | p \rangle $  is a pion  state with momentum $p$.
  Again,  on the lattice, we  take $z=z_3$ and 
   $\alpha=0$ to extract  the $p^\alpha {\cal M}(\nu, -z^2)$ 
  part of its decomposition over   Lorentz structures,
  and then form  the reduced Ioffe-time distribution ${\mathfrak M}(\nu, z_3^2)=
  {\cal M}(\nu,z_3^2)/  {\cal M}(0,z_3^2)$. 
  
 It can be shown \cite{Radyushkin:1983wh}    that, for all contributing  Feynman diagrams,   
 the   Fourier transform   of the pseudo-ITD  ${\cal M} (\nu, z_3^2)$ (and, hence, of  ${\mathfrak M}(\nu, z_3^2)$) 
   with respect to $\nu$ 
 has the $0\leq x \leq 1$ support.
 In other words, for any  $z_3^2$, we  may write 
   \begin{align}
 {\mathfrak M} (\nu, z_3^2) 
&   = 
 \int_{0}^1 dx 
 \, e^{i x \nu  } \,  \Phi (x, z_3^2)  \   ,
  \label{MPD}
\end{align}   
where $  \Phi (x, z_3^2) $ is the pion {\it pseudo-distribution amplitude} 
(pseudo-DA).  
Sometimes it is convenient to use the $(-z/2,z/2)$ endpoints instead of $(0,z)$.
Using  translation invariance, we get 
   \begin{align}
   \langle  0 | \bar \psi (-z_3/2) \,  \ldots 
\psi ( z_3/2)    | p \rangle  &=
e^{-i  \nu/2 }   {\cal M} (  \nu, z_3^2) \nn &  \equiv  \widetilde {\cal M} (\nu, z_3^2) \ . 
\label{Mtilde}
\end{align}

 To apply the general one-loop formula
(\ref{Oha}), we will need also a parameterization of the 
$\langle  0 | \bar \psi (uz_3) \, \ldots 
\psi (\bar v z)    | p \rangle $ matrix element.  
Again, by translation invariance, 
   \begin{align}
   \langle  0 | \bar \psi (uz_3) \,&  \ldots 
\psi (\bar v z_3)    | p \rangle  = 
e^{i u \nu}   {\cal M}_0 [(1-u-\varv)  \nu]  \nn 
&  =
 \int_{0}^1 dy 
 \, e^{i y \bar \varv \nu  + i \bar y u  \nu  } \,  \Phi_0 (y)  \   .
  \label{ubarv}
\end{align} 
This formula just says that the quark at $\bar v z_3$ has the $yp_3$ momentum, while 
that at $uz_3$ carries $\bar y p_3$.

\subsection{Structure of contributing terms} 
  
Let us start with  the evolution terms in Eq. (\ref{Oha}), i.e. with  those accompanied by $\ln (z_3^2)$ in Eq. (\ref{Oha}).
  Take first the ``vertex''  terms, i.e., those   containing $\delta (u) $ or $\delta (v)$.
  Then 
  \begin{align} 
 \,  & \int_0^1  du \int_0^1  d\varv  
    \langle  0 | \bar \psi (u z_3) \gamma^0 \gamma_5\psi (\bar \varv z_3)   | p \rangle
     \nn & \hspace{1cm} \times  \,   \left \{ \delta (u) \left[ \frac{\bar \varv }{\varv }\right ] _+
 + \delta (\varv) \left[ \frac{\bar u }{u }\right ] _+   \right \} \,\nn &
   =   \,\int_0^1  d\varv    \left[ \frac{\bar \varv }{\varv }\right ] _+
  \langle  0 |  \bar \psi (0) \gamma^0 \gamma_5  \psi (\bar \varv z_3) | p \rangle  
  \nn & 
 +  \int_0^1  du  \left[ \frac{\bar u }{u }\right ] _+  
  \, \langle  0 |  \bar \psi (u z_3) \gamma^0  \gamma_5 \psi ( z_3)  | p \rangle  
   \ .
    \label{evoda}
 \end{align}
Switching  to the $u+v =1-w$ notation, and using
translation invariance  we see that this is equal to 
  \begin{align} 
 \int_0^1  dw    \left[ \frac{w }{\bar w }  \right ] _+
    {\cal M} ( w \nu)
\left (1 +   \, e^{i \bar w \nu}  \right )  
    \ .
    \label{combo}
 \end{align}
 Transforming to the $\widetilde {\cal M} $-function using  Eq. (\ref{Mtilde}) in the form $ {\cal M} (  w \nu) = e^{i  w\nu/2 }  \widetilde {\cal M} (w\nu)$,
 we find  that Eq. (\ref{combo})  reduces to 
    \begin{align} 
  e^{i \nu/2} \
\,\int_0^1  du \,    \left[ \frac{2 w }{\bar w  }\right ] _+
  \widetilde  {\cal M} ( w \nu) 
\cos ( \bar w \nu/2  )  
    \ .
    \label{cos}
 \end{align}
 
 Take now the ``box'' term,    the
  integrand  of which is \mbox{$u,v$-independent} . Then 
   \begin{align} 
     \label{1da}
 \,  & \int_0^1  du \int_0^{1-v}   d\varv  
    \langle  0 | \bar \psi (u z_3) \gamma^0 \gamma_5\psi (\bar \varv z_3)   | p \rangle
    \nn & = 
 \,\int_0^1  du \int_0^{1-u}   d\varv \, 
 e^{i u \nu} {\cal M}_0 ((1-u-\varv) \nu)  
     \ .
 \end{align}
 Changing $u+v = 1-w, u =(1-w)\zeta$,  integrating over $\zeta$ 
 and switching to the $\widetilde {\cal M}$-function gives
  \begin{align} 
&   e^{i \nu/2} 
 \,\int_0^1  dw \,  
 \frac{ \sin (\overline{ w} \nu/2)}{\nu/2}  \widetilde  {\cal M}_0 (w\nu)  
     \ .
    \label{exch}
 \end{align}
Note that if we would calculate the correction to the function 
$\widetilde {\cal M} (\nu, z_3^2)  = e^{-i  \nu/2 }   {\cal M} (  \nu, z_3^2)$ rather than 
to $ {\cal M} (\nu, z_3^2)$, the overall factor of $ e^{i \nu/2} $ in Eqs. (\ref{cos}) and 
(\ref{exch})  would  be absent. 

\subsection{Matching} 

In a similar way,  one can derive formulas for other terms from Eq. (\ref{Oha}).
As, a result, we obtain an analog of Eq. (\ref{deltaM}), namely
  \begin{align} 
 \widetilde  {\mathcal M}(\nu,z_3^2)  & =  \Biggl [1-  \frac{\alpha_s}{2\pi} \, C_F  Z (z_3) \Biggr ]  \widetilde   {\cal M}_0(\nu) 
 \nn   &- 
\frac{\alpha_s}{2\pi} \, C_F 
\,\int_0^1  dw   \, \widetilde  {\mathcal  M}_0( w \nu) 
\Biggl  \{ \ln \left [ z_3^2 \mu_{\rm IR}^2\frac{ e^{2\gamma_E+1}}{4}  \right ]
\nn & \times \Biggr (
\Biggl  [\frac{2w }{1- w } \Biggr ]_+ \, \cos ( \bar w \nu/2  ) 
+  \frac{ \sin (\bar{ w} \nu/2)}{\nu/2} \Biggr )\nn & + 
   4    \Biggl [\frac{\ln (1-w) }{1-w } \Biggr ]_+ \cos ( \bar w \nu/2  )  
   -2  \frac{ \sin (\bar{ w} \nu/2)}{\nu/2} \Biggr \} \  .
\label{ITDmDA}
 \end{align}
 To form the reduced pseudo-ITD,
  \begin{align}
\widetilde {\mathfrak M} (\nu, z_3^2) \equiv \frac{\widetilde {\cal M} (\nu, z_3^2)}{\widetilde {\cal M} (0, z_3^2)} \  , 
 \label{redtilm}
\end{align}
 we need $ \widetilde  {\mathcal M}(0,z_3^2)$,
 which is given by 
   \begin{align} 
 \widetilde  {\mathcal M}&(0,z_3^2)   =  \Biggl [1-  \frac{\alpha_s}{2\pi} \, C_F  Z (z_3) \Biggr ]  \widetilde   {\cal M}_0(\nu) 
 \nn   &- 
\frac{\alpha_s}{2\pi} \, C_F 
\frac12  \widetilde  {\mathcal  M}_0( 0) 
\Biggl  \{ \ln \left [ z_3^2 \mu_{\rm IR}^2\frac{ e^{2\gamma_E+1}}{4}  \right ]
   -2   \Biggr \} \  .
\label{ITDmDAR}
 \end{align}
 Thus, the $ \sin (\bar{ w} \nu/2)/ (\nu/2)$ 
terms  present in  Eq. (\ref{ITDmDA})  
change into $ \sin (\bar{ w} \nu/2)/ (\nu/2) -\frac12 \delta( \bar w) $
in the expression for the reduced pseudo-ITD. 
This combination does not have a plus-prescription form, i.e., it differs from $[ \sin (\bar{ w} \nu/2)/ (\nu/2)]_+$, 
in contrast to 
 the PDF case, when $(1-w) -\frac12 \delta( \bar w) $
could be written as $(1-w)_+$. 

 However, just like in the PDF case, the $Z(z_3)$ term drops from the 
 ${\cal O} (\alpha_s)$ correction to the reduced pseudo-ITD.
 As a result, the matching condition in the pion DA  case is 
    \begin{align} 
&  \widetilde  {\mathcal I}(\nu,\mu^2)   =  \widetilde {\mathfrak M} (\nu, z_3^2)
 \nn &  +
\frac{\alpha_s}{2\pi} \, C_F 
\,\int_0^1  dw   \, \widetilde  {\mathfrak  M}( w \nu,z_3^2) 
\Biggl  \{ \ln \left [ z_3^2 \mu^2\frac{ e^{2\gamma_E+1}}{4}  \right ]
\nn & \times \Biggr (
\Biggl  [\frac{2w }{1- w } \Biggr ]_+ \, \cos ( \bar w \nu/2  ) 
+  \frac{ \sin (\bar{ w} \nu/2)}{\nu/2} -\frac12 \delta( \bar w)\Biggr )\nn & + 
   4    \Biggl [\frac{\ln (1-w) }{1-w } \Biggr ]_+ \cos ( \bar w \nu/2  )  
   -2  \frac{ \sin (\bar{ w} \nu/2)}{\nu/2}+ \delta( \bar w) \Biggr \} \  .
\label{ITDmDAf}
 \end{align}
 The ``tilde'' ITD $ \widetilde  {\cal I } (\nu, \mu^2)$ is related 
 to the light-cone pion DA $\Phi (x, \mu^2) $ by 
      \begin{align}
\widetilde  {\cal I } (\nu, \mu^2) 
&   = 
 \int_{0}^1 dx 
 \, e^{i (x-1/2) \nu  } \,  \Phi (x, \mu^2)  \   .
  \label{MPDa}
\end{align}   
   
   Again, the simplest way to extract $ \Phi (x, \mu^2)$ is to assume 
   some parameterization for it, like $N (x \bar x)^a$, and fit
   $a$ from the lattice data on $\widetilde  {\cal I } (\nu, \mu^2) $.
   
   Alternatively,  in analogy with Eq. (\ref{MtchI}), one may write $ \widetilde {\mathfrak M} (\nu, z_3^2)$ 
   in terms of  $  \Phi (x, \mu^2)$ and fit $\alpha_s$ and the parameters of 
   the $\Phi (x, \mu^2)$ model using the lattice data for $ \widetilde {\mathfrak M} (\nu, z_3^2)$. 
   The analog of $R(x \nu, z_3^2 \mu^2)$ of  Eq. (\ref{MtchI})  is also straightforward-calculable as a  closed-form expression, but 
   the result  is too long to present it here.
   
   A few more words about the  lattice implementation. While the function
   $ \widetilde {\mathfrak M} (\nu, z_3^2)$ is directly given by
   the matrix element of the operator with the $(-z_3/2,z_3/2)$ endpoints,
 a more practical  way to calculate  it is to  use the $(0,z_3)$ endpoints and multiply 
   the function ${\mathfrak M} (\nu, z_3^2)$ obtained in this way by the $e^{-i\nu/2}$
   factor to get  $ \widetilde {\mathfrak M} (\nu, z_3^2)$.
   The reason is that  $z_3/2$ on the lattice should be an integer  multiple
   of the lattice spacing $a$, say $z_3/2=na$.
   But then $z_3=2na$, i.e.,  the total separations $z_3$ given by an odd number 
   of spacings are   lost if one uses the $(-z_3/2,z_3/2)$ endpoints.

 \subsection{Checking the ERBL kernel}
 
 While the matching formula (\ref{ITDmDAf})  has a more
 involved form than that for PDFs, 
 the difference is basically the presence of sines and cosines 
of $w\nu/2$, which are smooth functions of $w$.

On the other hand, it is well known that the ERBL
(for Efremov-Radyushkin-Brodsky-Lepage \cite{Efremov:1978rn,Efremov:1979qk,Lepage:1980fj}) 
kernel $V(x,y)$ governing the evolution of the pion DA
is given by different functions for $x<y$ and $x>y$, i.e., it is
only piecewise smooth, with   
singularities like 
cusps, etc.,  for $x=y$.
 So, 
 one may wonder if Eq. (\ref{ITDmDA})  correctly reproduces the
 ERBL evolution equation
  \begin{align} 
\delta & \Phi  (x,z_3^2)   =  -  \frac{\alpha_s}{2\pi} \, C_F \ln (z_3^2) 
\int_0^1 dy \, V (x, y) \, \Phi_0 (y) \  + \ldots  \ 
 .
\label{Phiha}
 \end{align}
 
 Let us take first  the  ``vertex''  part corresponding 
 to \mbox{Eq. (\ref{combo})}  and write it in terms of the DA, 
   \begin{align} 
   \, &\int_0^1  dw    \left[ \frac{w }{\bar w  }\right ] _+
    {\cal M} ( w \nu)
\left (1 +   \, e^{i \bar w \nu}  \right ) \nn & = 
     \,\int_0^1  dw     \left[ \frac{w }{\bar w  }\right ] _+
    \int_{0}^1 dy 
 \, e^{i y w\nu  } \,  \Phi_0 (y)  
\left (1 +   \, e^{i  \bar w  \nu}  \right )  
    \ .
    \label{ker1}
 \end{align}
 Applying the  Fourier transformation 
    \begin{align}
  \Phi (x, z_3^2) = \frac1{2 \pi} \int_{-\infty}^\infty d \nu 
   \, e^{-i x \nu  } {\mathfrak M} (\nu,z_3^2) 
  \label{MDP}
\end{align}   
that  converts ${\cal M} (\nu)$ into  $\Phi (x)$, 
 gives  
  \begin{align} 
  &   
       \,\int_0^1  dw   \left[ \frac{w }{\bar w }\right ] _+ 
 \,  \left [  \delta (x   -wy) \, 
 +   \, \delta (- \bar x+ w \bar y )  \, 
 \right ] \nn & 
 =  \left[ \frac{y/x }{x-y}\right ] _+ 
 \,    \theta (y  <x) \, 
 +   \,   \left[ \frac{\bar y/\bar x  }{y-x  }\right ] _+ 
  \theta (y  >x) 
    \ , 
    \label{ERBL1}
 \end{align}
 which is a well-known part of the ERBL kernel $V(x,y)$ 
 (see, e.g., Ref. \cite{Lepage:1980fj}).
 As promised, it has different analytic forms  in the regions $x<y$ and $x>y$.
 
 For the ``box''   part given by  Eq. (\ref{1da}), we have 
   \begin{align} 
 &\,\int_0^1  du \int_0^{1-u}   d\varv \, 
 e^{i u \nu} {\cal M} ((1-u-\varv) \nu)  
 \nn & =
 \int_0^1  du \int_0^{1-u}   d\varv \, 
 e^{i u \nu}    \int_{0}^1 dy 
 \, e^{i y (1-u-v) \nu  } \,  \Phi (y)  
     \ .
          \label{boxker}
 \end{align}
 Applying the Fourier transform (\ref{MDP}) gives the
 remaining part  
  \begin{align} 
 &
\int_0^1  du \int_0^{1-u}   d\varv \, 
 \delta (x-y -u   + y(u+v))\nn & =
\frac{x}{y}  \theta (x\leq y) +\frac{\bar x }{\bar y}  \theta (x\leq y)
  \,  
     \ .
          \label{ERBL2}
 \end{align}
 of the ERBL kernel $V(x,y)$.  As a function of $x$, it
  is given by two straight lines intersecting at $x=y$, with a cusp at 
 this point. Its integral over $x$ gives 1/2, and  the \mbox{$-\frac12 \delta (1-w)$}  
 term in Eq. (\ref{ITDmDAf}) gives the contribution  $-\frac12 \delta (x-y)$ 
that provides  the plus-prescription for the kernel of Eq. (\ref{ERBL2}).

     \setcounter{equation}{0}
 
 \section{Matching for GPDs} 
 
\subsection{Definitions and kinematics}

For the pion, one may define the   light-cone GPDs  $H(x,\xi,t;\mu^2)$  \cite{Ji:1996ek} 
(see also Refs. \cite{Mueller:1998fv,Radyushkin:1997ki}) at a factorization scale $\mu$  by 
  \begin{align}
  \langle p_2 |&  \bar   \psi(-z/2) \gamma^\alpha \hat E(-z/2,z/2; A) \psi (z/2)|p_1 \rangle 
  \nn & = 2 {\cal P}^\alpha
  \int_{-1}^1 dx \, e^{-i x ({\cal P}z)} \, H(x,\xi,t;\mu^2) \ , 
\,  
 \
 \label{Hxi}
\end{align} 
where ${\cal P}= (p_1+p_2)/2$, the coordinate  $z$ has only the $z_-$ light-cone component
 and $\alpha=+$. 
The invariant momentum transfer is given by $t=(p_1-p_2)^2$.
The skewness variable $\xi$ is defined as 
  \begin{align}
  \xi = \frac{(p_1z)-(p_2z)}{(p_1z)+(p_2z)}
  \ . 
   \label{xi}
\end{align} 

For the nucleon, a similar definition  holds for the spin non-flip GPD 
$H(x,\xi,t;z^2)$, with $2{\cal P}^+$ substituted by $\bar u(p_2)\gamma^+ u(p_1)$. 

On the lattice, as discussed above,  it is more convenient 
to take the $ \bar \psi(0) \ldots  \psi (z)$  operator.
By translation invariance,
  \begin{align}
 & \langle p_2 |  \bar   \psi(0) \ldots  \psi (z)|p_1 \rangle 
  \nn & = e^{-i (p_1 z)/2 + i(p_2 z) /2}   \langle p_2 |  \bar   \psi(-z/2)\ldots  \psi (z/2)|p_1 \rangle \ .
\,  
 \
 \label{Hxisym}
\end{align} 

 To introduce pseudo-GPDs, we 
choose  $z=z_3$,    and take the average    momentum that is   also  oriented along 
the $z_3$  axis. Then 
$p_1=\{E_1,\Delta_\perp/2,P_1\}$ and 
\mbox{$p_2=\{E_2, - \Delta_\perp/2,P_2\}$.}  As a result, we  have two Ioffe-time invariants
\mbox{$\nu_1 \equiv -(p_1z) = P_1 z_3$}  and \mbox{$\nu_2 \equiv -(p_2z) = P_2 z_3$. }
Now we can define the double Ioffe-time pseudo-distribution $ {M} (\nu_1,\nu_2,t;z_3^2) $
  \begin{align}
  \langle p_2 | \bar   \psi(0) \gamma^0 \ldots  \psi (z_3)|p_1 \rangle = 2 {\cal P}^{0}  {M} (\nu_1,\nu_2,t;z_3^2)
\,   . 
 \
 \label{Mnn}
\end{align} 
The skewness variable  $\xi$ in this case is given by 
  \begin{align}
  \xi =& \frac{(p_1z_3) -(p_2 z_3)}{(p_1z_3) +(p_2 z_3)}  =\frac{P_1-P_2}{P_1+P_2}  \,
  =\frac{\nu_1-\nu_2}{\nu_1+\nu_2}
\,  .
 \
 \label{Zz1z2}
\end{align} 
Using this definition, we may write $P_1= (1+\xi)P$ and $P_2= (1-\xi)P$,
 where  
$P\equiv {\cal P}_3$. 
Denoting 
   \begin{align}\nu=\frac{\nu_1+\nu_2}{2} , 
   \end{align}  
 we  define the {\it generalized Ioffe-time pseudo-distribution} (pseudo-GITD) by  
\begin{align}
  &{M} (\nu_1,\nu_2,t;z_3^2) = {\cal M} (\nu, \xi ,t;z_3^2)\ , 
  \end{align} 
and parameterize it by  the {\it  pseudo-GPD} 
  \begin{align}
  &  {\cal M} (\nu, \xi,t;z_3^2)
 = e^{i \xi \nu } \,  \int_{-1}^1 dx \, e^{i x \nu } \, 
    {\cal H} \left (x,\xi,t;z_3^2\right )
\,  .
 \
 \label{MH}
\end{align} 
This formula tells us that the third momentum component  of the quark at the 
point $z_3$ is $(x+\xi)P$, as expected. 
The inverse transformation is given by 
  \begin{align}
 {\cal  H} \left (x,\xi,t;z_3^2\right ) =&\frac1{2\pi} \int_{-\infty}^\infty d\nu \, 
 e^{-i (x+\xi) \nu } \, {\cal M} (\nu, \xi,t;z_3^2)
   \,     
\,  .
 \
 \label{HM}
\end{align} 

Note that  originally we had two 
Ioffe-time parameters $\nu_1$ and $\nu_2$. However,  the Fourier representation  
(\ref{HM}) involves  integration over just one \mbox{$\nu$-parameter,} 
proportional to their sum. The difference $\nu_1-\nu_2$  is expressed in terms of  $\nu$ and    
 the skewness $\xi$ that  plays the role of a fixed parameter  like $t$ or $z_3^2$.

Just like in the pion DA case, it is convenient to introduce the ``tilde'' pseudo-GITD
  \begin{align}
  &  \widetilde {\cal M} (\nu, \xi,t;z_3^2)
 = e^{-i \xi \nu } \, {\cal M} (\nu, \xi,t;z_3^2)
\,  
 \
 \label{tildM}
\end{align} 
that is directly conjugate to the pseudo-GPD
  \begin{align}
  &  \widetilde {\cal M} (\nu, \xi,t;z_3^2)
 = \int_{-1}^1 dx \, e^{i x \nu } \, 
    {\cal H} \left (x,\xi,t;z_3^2\right )
\,  .
 \
 \label{Htxi}
\end{align} 

In deriving the matching relation, we will also need the representation 
 \begin{align}
  & \langle p_2 |  \bar  \psi( u z_3) \ldots  \psi (\bar \varv z_3)|p_1 \rangle \nn & =
e^{i (\nu_1-\nu_2)u } 
   {\cal M}_0 (\nu  (1-u-\varv) ,\xi)
   \nn & =
   e^{i2 \xi \nu u } 
   { \cal M}_0 (\nu (1-u-\varv) ,\xi  )
\,  
 \
 \label{Mzz}
\end{align}

 \subsection{Structure of contributing terms}

 Let us  now collect the  terms resulting from taking 
\mbox{ Eq. (\ref{Oha}) } between the $\langle p_2 | \ldots |p_1\rangle $ brackets. 
  Take first the ``vertex''  terms, i.e., those   containing $\delta (u) $ or $\delta (v)$.
 Proceeding as in the DA case,  we start with 
\begin{align} 
  &    \,\int_0^1  d\varv    \left[ \frac{\bar \varv }{\varv }\right ] _+
  \bar \psi (0) \gamma^\alpha \psi (\bar \varv z_3) \nn & 
 +  \int_0^1  du  \left[ \frac{\bar u }{u }\right ] _+  
  \, \bar \psi (u z_3) \gamma^\alpha \psi ( z_3) \ .
      \label{vertG}
 \end{align}
 Taking matrix elements we arrive at 
     \begin{align}
   &   \,\int_0^1  d\varv    \left[ \frac{\bar \varv }{\varv }\right ] _+
 {\cal M}_0 (\nu (1-\varv) , \xi ) \nn &
 +  \int_0^1  du  \left[ \frac{\bar u }{u }\right ] _+  
  \,
  e^{i2 \xi \nu u } 
   {\cal M}_0 (\nu (1-u) ,\xi) \nn &=
    \int_0^1  dw  \left[ \frac{w }{\bar w}\right ] _+  
  \,
 (1+ e^{i2 \xi \nu \bar w  } )
   {\cal M}_0 (w \nu  ,\xi)
      \ .
    \label{vertG2}
 \end{align}
 Switching to the $\widetilde {\cal M}_0$-function, we transform this expression into 
     \begin{align}
 & e^{i \xi \nu  } 
      \int_0^1  dw \,  \left[ \frac{2w }{1-w}\right ] _+  
  \,
 \cos (\xi \nu\bar  w )  
   \widetilde {\cal M}_0 (w \nu  ,\xi) 
      \ .
    \label{vertGt}
 \end{align}
 Again, the overall $ e^{i \xi \nu  } $ factor tells us that this is a correction to the 
 ${\cal M}$-function written in terms of the \mbox{$\widetilde {\cal M}_0$-functions.}
 
 To check what kind of evolution kernel we have now,
  write the last line of Eq. (\ref{vertG2}) in terms of the GPD.
 This gives
    \begin{align}
& 
    \int_0^1  dw  \left[ \frac{w }{\bar w}\right ] _+  
  \,
 (1+ e^{i2 \xi \nu \bar w  } )
   {\cal M}_0 (w \nu  ,\xi) \nn = &
      \int_0^1  dw  \left[ \frac{w }{\bar w}\right ] _+  
  \,
 (1+ e^{i2 \xi \nu \bar w  } )
\,  \int_{-1}^1 dy \, e^{i (y+\xi)  \nu w } \, 
    H_0 \left (y,\xi\right ) \ .
     \end{align}
   Applying the Fourier transformation  (\ref{HM}) that converts ${\cal M}$ into $H$, we
   get the following representation for the  ``vertex''  part  of the GPD evolution  kernel
    \begin{align}
   K_v(x,y;\xi) =  & \int_0^1  dw  \left[ \frac{w }{\bar w }\right ] _+  
  \,
 \,  \int_{-1}^1 dy \, \,[  \delta((y+\xi)  w -(x + \xi))  \nn & + \delta ( (y-\xi) w -(x-\xi) )  ]
      \ .
    \label{Gker}
 \end{align}
 It is easy to check that, for $\xi=0$, this expression gives the ``vertex'' part of the AP kernel, 
 while for $\xi=1$ it gives the ``vertex''  part (\ref{ERBL1}) of the ERBL kernel.

 Consider  now the ``box'' term which has the 
  \mbox{$u,v$-independent}  integrand. Then we deal with
   \begin{align} 
 \,  &  \,\int_0^1  du \int_0^{1-u}   d\varv \, 
 e^{i2 \xi \nu u }   {\cal M}_0 ((1-u-\varv) \nu)  
     \ .
 \end{align}
  Changing $u+v = 1-w, u =(1-w)\zeta$,  integrating over $\zeta$ 
 and switching to the $\widetilde {\cal M}$-function gives
  \begin{align} 
&   e^{i \xi \nu} 
 \,\int_0^1  dw \,  
 \frac{ \sin (\bar w \xi  \nu)}{\xi \nu}  \widetilde  {\cal M}_0 (w\nu)  
     \ .
    \label{exchG}
 \end{align}
 Just like in Eq. (\ref{vertGt}), 
 we have here an overall factor of $ e^{i \xi \nu} $, as expected. 

Further steps go absolutely in parallel with the derivation of
the matching relation for the pion DA.
Skipping these  steps, we present here  the final result 
\begin{align} 
&  \widetilde  {\mathcal I}(\nu,\xi,t,\mu^2)   =  \widetilde {\mathfrak M} (\nu, \xi, t, z_3^2)
 \nn &    +
\frac{\alpha_s}{2\pi} \, C_F 
\,\int_0^1  dw   \, \widetilde  {\mathfrak  M}( w \nu, \xi, t, z_3^2) 
\Biggl  \{ \ln \left [ z_3^2 \mu^2\frac{ e^{2\gamma_E+1}}{4}  \right ]
\nn & \times \Biggr (
\Biggl  [\frac{2w }{1- w } \Biggr ]_+ \, \cos ( \bar w \xi \nu  ) 
+  \frac{ \sin (\bar{ w} \xi \nu)}{\xi \nu} -\frac12 \delta( \bar w)\Biggr )\nn & + 
   4    \Biggl [\frac{\ln (1-w) }{1-w } \Biggr ]_+ \cos ( \bar w \xi \nu  )  
   -2  \frac{ \sin (\bar{ w} \xi  \nu)}{\xi \nu}+ \delta( \bar w) \Biggr \} \  
\label{ITDmGPD}
 \end{align}
 that gives the light-cone  GITD 
  \begin{align} 
&  \widetilde  {\mathcal I}(\nu,\xi,t,\mu^2)   =   \int_{-1}^1 dx\, e^{i x \nu} H(x,\xi,t;\mu^2)
\label{LCITDGPD}
 \end{align}
 in terms of 
 the reduced pseudo-GITD
   \begin{align}
\widetilde {\mathfrak M} (\nu, \xi,t, z_3^2) \equiv \frac{\widetilde 
{\cal M} (\nu, \xi, t, z_3^2)}{\widetilde {\cal M} (0,0,0, z_3^2)} \  .
 \label{redITDGPD}
\end{align}

To extract $H(x,\xi,t;\mu^2)$, we again propose to take some parameterization 
for it, and then fit the parameters using the lattice data on $\widetilde {\mathfrak M} (\nu, \xi,t, z_3^2) $.
Doing this, one should  keep in mind  that the GPD has  a non-trivial {\it polynomiality} property \cite{Ji:1996ek,Mueller:1998fv,Radyushkin:1997ki}.   It amounts to the requirement  
that, in the non-singlet case,  its $x^N$ moment should be a polynomial of the  $N$th degree in $\xi$. 
A possible  way to satisfy it is to use the {\it double distribution Ansatz} \cite{Radyushkin:1998es}. 

An equivalent alternative strategy, just like in the PDF and DA cases,
is to start with  the matching relation between  the  reduced pseudo-GITD $\widetilde {\mathfrak M} (\nu, \xi,t, z_3^2)$ 
and  the light-cone  GITD 
 $ \widetilde  {\mathcal I}(\nu,\xi,t,\mu^2) $ written in terms of $H(x,\xi,t;\mu^2)$ through Eq. (\ref{LCITDGPD}),
 and fit the parameters of  $H(x,\xi,t;\mu^2)$  from the lattice data on $\widetilde {\mathfrak M} (\nu, \xi,t, z_3^2)$. 

\subsection{Remarks on lattice implementation}

Just like in the pion DA case, on the lattice  it is more practical  
to measure matrix elements $M (\nu_1,\nu_2,t;z_3^2)$   of  the operators 
with the  $(0,z)$ endpoints, and then  to multiply them   by $e^{-i \xi \nu } = 
e^{-i (\nu_1-\nu_2)/2 } $  to convert the  result into the $\widetilde {\cal M}(\nu, \xi,t,z_3^2)$ functions 
corresponding to the $(-z/2,z/2)$ endpoints. 

Furthermore, on the lattice, the measurements will be done 
on a discrete  set of  values for coordinates $z_3 = n_z a$ and longitudinal momenta $P_1= 2\pi N_1 /L$,   
$P_2= 2\pi N_2 /L$, where $L=n a$ is the lattice size in the $z_3$ direction. 
Thus, possible values of the Ioffe-time parameters  are limited to 
discrete sets $\nu_1=2\pi n_z N_1/n$ and  $\nu_2=2\pi n_z N_2/n$.
Correspondingly, possible values for skewness are given by a set of 
rational numbers 
   \begin{align}
\xi = \frac{P_1-P_2}{P_1+P_2} =\frac{N_1-N_2}{N_1+N_2} \ . 
\label{xipn}
\end{align} 
In particular, changing $N_1$ and $N_2$ from 0 to 6,
gives  13 possible values for $\xi$ ranging from 0 to 1 and 
rather well representing the whole $0\leq \xi \leq 1$ segment.

However, varying the   value of $\xi$  also changes the value of the momentum transfer $t$.
 Namely, taking purely longitudinal momenta 
   \begin{align}
p_1 =& \{E_1, 0_\perp, P_1\} = \{E_1, 0_\perp,  (1+\xi )P\} %\equiv B(z,p)
\nn   
 p_2 =& \{E_2, 0_\perp, P_2\} = \{E_2, 0_\perp,  (1-\xi )P\}
\,  
 \
 \label{p1p2}
\end{align}  
with 
   \begin{align}
E_1 =&\sqrt{ M^2+   P_1^2} %\equiv B(z,p)
 \ \ \ , \ \ \   
 E_2 =\sqrt{ M^2+   P_2^2} 
\,  .
 \
 \label{E1E2}
\end{align} 
we get
\begin{align}
 t  =&  -
 \frac{2{M^2} (  P_1-   P_2)^2}{  M^2+ P_1 P_2   +
 \sqrt{  M^2+ P_1^2} \sqrt{M^2 + P_2^2}}\nn & 
 \equiv {t }_0(P_1,P_2,M)
\,  ,
 \
 \label{tP1P2}
\end{align} 
or, in the $\{\xi, P\}$ variables,
\begin{align}
{t_0 }= & - \frac{8\xi^2 {M^2}  }{   1-\xi^2+\frac{M^2}{P^2}  +
 \sqrt{  ( 1-\xi^2+\frac{M^2}{P^2})^2  + 4\xi^2 \frac{M^2}{P^2}}}
\,  .
 \
 \label{tmin}
\end{align} 
When $M^2/P^2$ is small (this is  not a very realistic situation for the nucleon, but still), we have 
\begin{align}
t_0   \approx  & - \frac{4\xi^2 {M^2}  }{   1-\xi^2}
\,  .
 \
 \label{tapprox}
\end{align} 
For small $\xi$, we can approximate
\begin{align}
t_0  \approx  & - \frac{4\xi^2 {M^2}  }{   1+M^2/P^2}
\,  .
 \
 \label{t0}
\end{align} 
In these formulas, $t_0 $ increases when $\xi$ increases. In any case, 
 this  value of $t$ is $\xi$-dependent, while 
we need  to extract GPDs as functions of $x$  for   fixed   $\xi$  and $t$.

To solve this problem, one may 
 add a transverse component $\Delta_\perp$ to the momentum 
transfer, i.e., to use \mbox{$p_1=\{E_1,\Delta_\perp/2,P_1\}$}  and $p_2=\{E_2,-\Delta_\perp/2,P_2\}$.
Then
 \begin{align}
&t = 
2M^2+ 2 P_1 P_2-\Delta_\perp^2/2  \nn & - 2 \sqrt{ M^2+  P_1^2+\Delta_\perp^2/4}  \sqrt{ M^2+  P_2^2+\Delta_\perp^2/4} 
\,  .
 \
 \label{tDel}
\end{align}

Again, on the lattice,  we will have a discrete set of possible
$\Delta_\perp^2$ values.  As a result,    it is   impossible 
to arrange exactly the same value of $t$ for different values of $\xi$.
A more modest   goal is to collect  a set of data 
with close values of $t$, and then make interpolation to a chosen \mbox{$t$-value.  }

 Another strategy is to choose first some particular values of $P_1$ and $P_2$.
This  fixes the value of $\xi$.  Then we take  several different values of $\Delta_\perp$
 to change $t$. That will give the \mbox{$t$-dependence}  for fixed $\xi$ and $\nu$.
 After this, changing $z_3$, we will change $\nu$ leaving $\xi$ and $t$ unchanged.
 Then, using  the matching conditions to convert the $\nu$-dependence into the $x$-dependence,
 we will end up with  $H(x,\xi,t;\mu^2)$ for a fixed $\xi$ as a function  of $x$ and $t$.

\section{ Summary.}

In this  paper, we have derived the matching relations for 
the pion distribution amplitude and non-singlet generalized parton
distributions that connect them with their off-the-light-cone counterparts,
the pseudo-DA and pseudo-GPDs. 
The latter may be calculated in lattice simulations,
and the matching relations are crucial in  converting them 
into the experimentally measurable (in principle) light-cone parton distributions.
 
 The main feature of our derivations is that
 we start with a universal expression for the one-loop 
 correction in an operator form.
 Then we show how this universal expression
 produces particular  matching conditions for ITDs related to 
 different parton distributions.
 In fact, these different matching relations have a rather similar 
 structure.  
 Also, these relations are much simpler than the matching 
 relations for quasi-PDFs, quasi-DAs and quasi-GPDs given in Refs.
  \mbox{\cite{Ji:2013dva,Xiong:2013bka,Ji:2015jwa,Izubuchi:2018srq,Wang:2019tgg, Ji:2013dva,Xiong:2013bka,Ji:2015jwa,Izubuchi:2018srq,Wang:2019tgg}}.
 
 The matching relations for the  pseudo-PDFs have been already
 used in  lattice calculations \cite{Orginos:2017kos,Radyushkin:2018cvn,Karpie:2018zaz,Joo:2019jct,Joo:2019bzr},
 while these for the pion DA and GPDs will be used in the ongoing lattice calculations.

\acknowledgements

I thank  K. Orginos  and R. Edwards   for  their 
 interest in  this investigation and discussions.   
This work is supported by Jefferson Science Associates,
 LLC under  U.S. DOE Contract \mbox{ \#DE-AC05-06OR23177} 
 and by U.S. DOE Grant \#DE-FG02-97ER41028.


\begin{thebibliography}{10}
 
 
%\cite{Cichy:2018mum}
\bibitem{Cichy:2018mum} 
  K.~Cichy and M.~Constantinou,
  %``A guide to light-cone PDFs from Lattice QCD: an overview of approaches, techniques and results,''
  Adv.\ High Energy Phys.\  {\bf 2019}, 3036904 (2019)
  %doi10.1155/2019/3036904
  %[arXiv:1811.07248 [hep-lat]].
  %%CITATION = %doi10.1155/2019/3036904;%%
  %15 citations counted in INSPIRE as of 14 Sep 2019
 

\bibitem{Ji:2013dva} 
  X.~Ji,
  %``Parton Physics on a Euclidean Lattice,''
  Phys.\ Rev.\ Lett.\  {\bf 110}, 262002 (2013)
  %doi10.1103/PhysRevLett.110.262002
  %[arXiv:1305.1539 [hep-ph]].
  %%CITATION = %doi10.1103/PhysRevLett.110.262002;%%
  %260 citations counted in INSPIRE as of 14 Sep 2019

%\cite{Ji:2014gla}
\bibitem{Ji:2014gla} 
  X.~Ji,
  %``Parton Physics from Large-Momentum Effective Field Theory,''
  Sci.\ China Phys.\ Mech.\ Astron.\  {\bf 57}, 1407 (2014)
  %doi10.1007/s11433-014-5492-3
  %[arXiv:1404.6680 [hep-ph]].
  %%CITATION = %doi10.1007/s11433-014-5492-3;%%
  %97 citations counted in INSPIRE as of 14 Sep 2019

% \cite{Detmold:2005gg}
\bibitem{Detmold:2005gg} 
  W.~Detmold and C.~J.~D.~Lin,
  %``Deep-inelastic scattering and the operator product expansion in lattice QCD,''
  Phys.\ Rev.\ D {\bf 73}, 014501 (2006)
  %doi10.1103/PhysRevD.73.014501
  %[hep-lat/0507007].
  %%CITATION = %doi10.1103/PhysRevD.73.014501;%%
  %61 citations counted in INSPIRE as of 15 Sep 2019 
  
  %\cite{Braun:2007wv}
\bibitem{Braun:2007wv} 
  V.~Braun and D.~M\"uller,
  %``Exclusive processes in position space and the pion distribution amplitude,''
  Eur.\ Phys.\ J.\ C {\bf 55}, 349 (2008)
  %doi10.1140/epjc/s10052-008-0608-4
  %[arXiv:0709.1348 [hep-ph]].
  %%CITATION = %doi10.1140/epjc/s10052-008-0608-4;%%
  %54 citations counted in INSPIRE as of 16 Sep 2019
  
  %\cite{Ma:2014jla}
\bibitem{Ma:2014jla} 
  Y.~Q.~Ma and J.~W.~Qiu,
  %``Extracting Parton Distribution Functions from Lattice QCD Calculations,''
  Phys.\ Rev.\ D {\bf 98}, no. 7, 074021 (2018)
  %doi10.1103/PhysRevD.98.074021
  %[arXiv:1404.6860 [hep-ph]].
  %%CITATION = %doi10.1103/PhysRevD.98.074021;%%
  %100 citations counted in INSPIRE as of 16 Sep 2019
  
  %\cite{Ma:2017pxb}
\bibitem{Ma:2017pxb} 
  Y.~Q.~Ma and J.~W.~Qiu,
  %``Exploring Partonic Structure of Hadrons Using ab initio Lattice QCD Calculations,''
  Phys.\ Rev.\ Lett.\  {\bf 120}, no. 2, 022003 (2018)
  %doi10.1103/PhysRevLett.120.022003
  %[arXiv:1709.03018 [hep-ph]].
  %%CITATION = %doi10.1103/PhysRevLett.120.022003;%%
  %67 citations counted in INSPIRE as of 16 Sep 2019







%\cite{Radyushkin:2017cyf}
\bibitem{Radyushkin:2017cyf} 
  A.~V.~Radyushkin,
  %``Quasi-parton distribution functions, momentum distributions, and pseudo-parton distribution functions,''
  Phys.\ Rev.\ D {\bf 96}, no. 3, 034025 (2017)
  %doi10.1103/PhysRevD.96.034025
  %[arXiv:1705.01488 [hep-ph]].
  %%CITATION = %doi10.1103/PhysRevD.96.034025;%%
  %82 citations counted in INSPIRE as of 16 Sep 2019
  
  %\cite{Radyushkin:2017sfi}
\bibitem{Radyushkin:2017sfi} 
  A.~Radyushkin,
  %``Quasi-PDFs and pseudo-PDFs,''
  PoS QCDEV {\bf 2017}, 021 (2017),
  %doi10.22323/1.308.0021
  [arXiv:1711.06031 [hep-ph]].
  %%CITATION = %doi10.22323/1.308.0021;%%
  %3 citations counted in INSPIRE as of 16 Sep 2019
  
  %\cite{Orginos:2017kos}
\bibitem{Orginos:2017kos} 
  K.~Orginos, A.~Radyushkin, J.~Karpie and S.~Zafeiropoulos,
  %``Lattice QCD exploration of parton pseudo-distribution functions,''
  Phys.\ Rev.\ D {\bf 96}, no. 9, 094503 (2017)
  %doi10.1103/PhysRevD.96.094503
  %[arXiv:1706.05373 [hep-ph]].
  %%CITATION = %doi10.1103/PhysRevD.96.094503;%%
  %78 citations counted in INSPIRE as of 16 Sep 2019
  
  
  %\cite{Xiong:2013bka}
\bibitem{Xiong:2013bka} 
  X.~Xiong, X.~Ji, J.~H.~Zhang and Y.~Zhao,
  %``One-loop matching for parton distributions: Nonsinglet case,''
  Phys.\ Rev.\ D {\bf 90}, no. 1, 014051 (2014)
  %doi10.1103/PhysRevD.90.014051
  %[arXiv:1310.7471 [hep-ph]].
  %%CITATION = %doi10.1103/PhysRevD.90.014051;%%
  %116 citations counted in INSPIRE as of 16 Sep 2019
  
  %\cite{Ji:2015jwa}
\bibitem{Ji:2015jwa} 
  X.~Ji and J.~H.~Zhang,
  %``Renormalization of quasiparton distribution,''
  Phys.\ Rev.\ D {\bf 92}, 034006 (2015)
  %doi10.1103/PhysRevD.92.034006
  %[arXiv:1505.07699 [hep-ph]].
  %%CITATION = %doi10.1103/PhysRevD.92.034006;%%
  %60 citations counted in INSPIRE as of 16 Sep 2019
  
  %\cite{Izubuchi:2018srq}
\bibitem{Izubuchi:2018srq} 
  T.~Izubuchi, X.~Ji, L.~Jin, I.~W.~Stewart and Y.~Zhao,
  %``Factorization Theorem Relating Euclidean and Light-Cone Parton Distributions,''
  Phys.\ Rev.\ D {\bf 98}, no. 5, 056004 (2018)
  %doi10.1103/PhysRevD.98.056004
  %[arXiv:1801.03917 [hep-ph]].
  %%CITATION = %doi10.1103/PhysRevD.98.056004;%%
  %43 citations counted in INSPIRE as of 16 Sep 2019
  
  %\cite{Wang:2019tgg}
\bibitem{Wang:2019tgg} 
  W.~Wang, J.~H.~Zhang, S.~Zhao and R.~Zhu,
  %``A Complete Matching for Quasi-distribution Functions in Large Momentum Effective Theory,''
  arXiv:1904.00978 [hep-ph]
  %%CITATION = ARXIV:1904.00978;%%
  %3 citations counted in INSPIRE as of 16 Sep 2019
  
  
  %\cite{Ji:2015qla}
\bibitem{Ji:2015qla} 
  X.~Ji, A.~Sch{\"a}fer, X.~Xiong and J.~H.~Zhang,
  %``One-Loop Matching for Generalized Parton Distributions,''
  Phys.\ Rev.\ D {\bf 92}, 014039 (2015)
  %doi10.1103/PhysRevD.92.014039
  %[arXiv:1506.00248 [hep-ph]].
  %%CITATION = %doi10.1103/PhysRevD.92.014039;%%
  %48 citations counted in INSPIRE as of 16 Sep 2019

%\cite{Xiong:2015nua}
\bibitem{Xiong:2015nua} 
  X.~Xiong and J.~H.~Zhang,
  %``One-loop matching for transversity generalized parton distribution,''
  Phys.\ Rev.\ D {\bf 92}, no. 5, 054037 (2015)
  %doi10.1103/PhysRevD.92.054037
  %[arXiv:1509.08016 [hep-ph]].
  %%CITATION = %doi10.1103/PhysRevD.92.054037;%%
  %37 citations counted in INSPIRE as of 16 Sep 2019
  
  %\cite{Liu:2019urm}
\bibitem{Liu:2019urm} 
  Y.~S.~Liu, W.~Wang, J.~Xu, Q.~A.~Zhang, J.~H.~Zhang, S.~Zhao and Y.~Zhao,
  %``Matching generalized parton quasidistributions in the RI/MOM scheme,''
  Phys.\ Rev.\ D {\bf 100}, no. 3, 034006 (2019)
  %doi10.1103/PhysRevD.100.034006
  %[arXiv:1902.00307 [hep-ph]].
  %%CITATION = %doi10.1103/PhysRevD.100.034006;%%
  %3 citations counted in INSPIRE as of 16 Sep 2019
  
  
  %\cite{Ji:2017rah}
\bibitem{Ji:2017rah} 
  X.~Ji, J.~H.~Zhang and Y.~Zhao,
  %``More On Large-Momentum Effective Theory Approach to Parton Physics,''
  Nucl.\ Phys.\ B {\bf 924}, 366 (2017)
  %doi10.1016/j.nuclphysb.2017.09.001
  %[arXiv:1706.07416 [hep-ph]].
  %%CITATION = %doi10.1016/j.nuclphysb.2017.09.001;%%
  %45 citations counted in INSPIRE as of 16 Sep 2019
  
  %\cite{Radyushkin:2017lvu}
\bibitem{Radyushkin:2017lvu} 
  A.~V.~Radyushkin,
  %``Quark pseudodistributions at short distances,''
  Phys.\ Lett.\ B {\bf 781}, 433 (2018)
  %doi10.1016/j.physletb.2018.04.023
  %[arXiv:1710.08813 [hep-ph]].
  %%CITATION = %doi10.1016/j.physletb.2018.04.023;%%
  %15 citations counted in INSPIRE as of 16 Sep 2019
  
  %\cite{Radyushkin:2018cvn}
\bibitem{Radyushkin:2018cvn} 
  A.~Radyushkin,
  %``One-loop evolution of parton pseudo-distribution functions on the lattice,''
  Phys.\ Rev.\ D {\bf 98}, no. 1, 014019 (2018)
  %doi10.1103/PhysRevD.98.014019
  %[arXiv:1801.02427 [hep-ph]].
  %%CITATION = %doi10.1103/PhysRevD.98.014019;%%
  %17 citations counted in INSPIRE as of 16 Sep 2019
  
  %\cite{Zhang:2018ggy}
\bibitem{Zhang:2018ggy} 
  J.~H.~Zhang, J.~W.~Chen and C.~Monahan,
  %``Parton distribution functions from reduced Ioffe-time distributions,''
  Phys.\ Rev.\ D {\bf 97}, no. 7, 074508 (2018)
  %doi10.1103/PhysRevD.97.074508
  %[arXiv:1801.03023 [hep-ph]].
  %%CITATION = %doi10.1103/PhysRevD.97.074508;%%
  %18 citations counted in INSPIRE as of 16 Sep 2019
  
  %\cite{Balitsky:1987bk}
\bibitem{Balitsky:1987bk} 
  I.~I.~Balitsky and V.~M.~Braun,
  %``Evolution Equations for QCD String Operators,''
  Nucl.\ Phys.\ B {\bf 311}, 541 (1989)
  %%doi10.1016/0550-3213(89)90168-5
  %%CITATION = %%doi10.1016/0550-3213(89)90168-5;%%
  %456 citations counted in INSPIRE as of 23 Jan 2018
  

%\cite{Ioffe:1969kf}
\bibitem{Ioffe:1969kf} 
  B.~L.~Ioffe,
  %``Space-time picture of photon and neutrino scattering and electroproduction cross-section asymptotics,''
  Phys.\ Lett.\  {\bf 30B}, 123 (1969)
  %%doi10.1016/0370-2693(69)90415-8
  %%CITATION = %%doi10.1016/0370-2693(69)90415-8;%%
  %264 citations counted in INSPIRE as of 23 Jan 2018


%\cite{Radyushkin:2016hsy}
\bibitem{Radyushkin:2016hsy} 
  A.~Radyushkin,
  %``Nonperturbative Evolution of Parton Quasi-Distributions,''
  Phys.\ Lett.\ B {\bf 767}, 314 (2017)
  %doi10.1016/j.physletb.2017.02.019
  %[arXiv:1612.05170 [hep-ph]].
  %%CITATION = %doi10.1016/j.physletb.2017.02.019;%%
  %57 citations counted in INSPIRE as of 16 Sep 2019
  
  
%\cite{Altarelli:1977zs}
\bibitem{Altarelli:1977zs} 
  G.~Altarelli and G.~Parisi,
  %``Asymptotic Freedom in Parton Language,''
  Nucl.\ Phys.\ B {\bf 126}, 298 (1977)
  %%doi10.1016/0550-3213(77)90384-4
  %%CITATION = %%doi10.1016/0550-3213(77)90384-4;%%
  %6223 citations counted in INSPIRE as of 23 Jan 2018

%\cite{Gribov:1972ri}
\bibitem{Gribov:1972ri} 
  V.~N.~Gribov and L.~N.~Lipatov,
  %``Deep inelastic e p scattering in perturbation theory,''
  Sov.\ J.\ Nucl.\ Phys.\  {\bf 15}, 438 (1972)
  [Yad.\ Fiz.\  {\bf 15}, 781 (1972)].
  %%CITATION = SJNCA,15,438;%%
  %4066 citations counted in INSPIRE as of 16 Sep 2019
  
  %\cite{Dokshitzer:1977sg}
\bibitem{Dokshitzer:1977sg} 
  Y.~L.~Dokshitzer,
  %``Calculation of the Structure Functions for Deep Inelastic Scattering and e+ e- Annihilation by Perturbation Theory in Quantum Chromodynamics.,''
  Sov.\ Phys.\ JETP {\bf 46}, 641 (1977)
  [Zh.\ Eksp.\ Teor.\ Fiz.\  {\bf 73}, 1216 (1977)].
  %%CITATION = SPHJA,46,641;%%
  %3673 citations counted in INSPIRE as of 16 Sep 2019
  
  
%\cite{Ji:2017oey}
\bibitem{Ji:2017oey}
  X.~Ji, J.~H.~Zhang and Y.~Zhao,
  %``Renormalization in Large Momentum Effective Theory of Parton Physics,''
  Phys.\ Rev.\ Lett.\  {\bf 120} (2018) no.11,  112001
  %doi:10.1103/PhysRevLett.120.112001
  %[arXiv:1706.08962 [hep-ph]].
  %%CITATION = %doi:10.1103/PhysRevLett.120.112001;%%
  %37 citations counted in INSPIRE as of 18 Jul 2018


%\cite{Ishikawa:2017faj}
\bibitem{Ishikawa:2017faj} 
  T.~Ishikawa, Y.~Q.~Ma, J.~W.~Qiu and S.~Yoshida,
  %``Renormalizability of quasiparton distribution functions,''
  Phys.\ Rev.\ D {\bf 96}, no. 9, 094019 (2017)
  %%doi10.1103/PhysRevD.96.094019
 % %[arXiv:1707.03107 [hep-ph]].
  %%CITATION = %%doi10.1103/PhysRevD.96.094019;%%
  %25 citations counted in INSPIRE as of 23 Jan 2018








%\cite{Green:2017xeu}
\bibitem{Green:2017xeu}
  J.~Green, K.~Jansen and F.~Steffens,
  %``Nonperturbative renormalization of nonlocal quark bilinears for quasi-PDFs on the lattice using an auxiliary field,''
  Phys.\ Rev.\ Lett.\  {\bf 121} (2018) no.2,  022004
  %doi:10.1103/PhysRevLett.121.022004
  %[arXiv:1707.07152 [hep-lat]].
  %%CITATION = %doi:10.1103/PhysRevLett.121.022004;%%
  %33 citations counted in INSPIRE as of 18 Jul 2018
  



%\cite{Braun:1994jq}
\bibitem{Braun:1994jq} 
  V.~Braun, P.~Gornicki and L.~Mankiewicz,
  %``Ioffe - time distributions instead of parton momentum distributions in description of deep inelastic scattering,''
  Phys.\ Rev.\ D {\bf 51}, 6036 (1995)
  %%doi10.1103/PhysRevD.51.6036
  %[hep-ph/9410318].
  %%CITATION = %%doi10.1103/PhysRevD.51.6036;%%
  %43 citations counted in INSPIRE as of 23 Jan 2018
  
  %\cite{Karpie:2019eiq}
\bibitem{Karpie:2019eiq} 
  J.~Karpie, K.~Orginos, A.~Rothkopf and S.~Zafeiropoulos,
  %``Reconstructing parton distribution functions from Ioffe time data: from Bayesian methods to Neural Networks,''
  JHEP {\bf 1904}, 057 (2019)
%  doi:10.1007/JHEP04(2019)057
%  [arXiv:1901.05408 [hep-lat]].
  %%CITATION = doi:10.1007/JHEP04(2019)057;%%
  %8 citations counted in INSPIRE as of 16 Sep 2019

%\cite{Accardi:2016qay}
\bibitem{Accardi:2016qay} 
  A.~Accardi, L.~T.~Brady, W.~Melnitchouk, J.~F.~Owens and N.~Sato,
  %``Constraints on large-$x$ parton distributions from new weak boson production and deep-inelastic scattering data,''
  Phys.\ Rev.\ D {\bf 93}, no. 11, 114017 (2016)
  %doi:10.1103/PhysRevD.93.114017
 % [arXiv:1602.03154 [hep-ph]].
  %%CITATION = doi:10.1103/PhysRevD.93.114017;%%
  %129 citations counted in INSPIRE as of 16 Sep 2019
  
%\cite{Cichy:2019ebf}
\bibitem{Cichy:2019ebf} 
  K.~Cichy, L.~Del Debbio and T.~Giani,
  %``Parton distributions from lattice data: the nonsinglet case,''
  arXiv:1907.06037 [hep-ph].
  %%CITATION = ARXIV:1907.06037;%%
  %3 citations counted in INSPIRE as of 21 Sep 2019
  
   %\cite{Radyushkin:2018nbf}
\bibitem{Radyushkin:2018nbf} 
  A.~V.~Radyushkin,
  %``Structure of parton quasi-distributions and their moments,''
  Phys.\ Lett.\ B {\bf 788}, 380 (2019)
  doi:10.1016/j.physletb.2018.11.047
  [arXiv:1807.07509 [hep-ph]].
  %%CITATION = doi:10.1016/j.physletb.2018.11.047;%%
  %18 citations counted in INSPIRE as of 23 Sep 2019

%\cite{Radyushkin:1977gp}
\bibitem{Radyushkin:1977gp} 
  A.~V.~Radyushkin, JINR-Dubna preprint P2-10717 (1977), 
  %``Deep Elastic Processes of Composite Particles in Field Theory and Asymptotic Freedom,''
  %Submitted to: Phys.Lett.
  arXiv: hep-ph/0410276
  %%CITATION = HEP-PH/0410276;%%
  %77 citations counted in INSPIRE as of 16 Sep 2019
  
  %\cite{Lepage:1980fj}
\bibitem{Lepage:1980fj} 
  G.~P.~Lepage and S.~J.~Brodsky,
  %``Exclusive Processes in Perturbative Quantum Chromodynamics,''
  Phys.\ Rev.\ D {\bf 22}, 2157 (1980)
 % doi:10.1103/PhysRevD.22.2157
  %%CITATION = doi:10.1103/PhysRevD.22.2157;%%
  %3497 citations counted in INSPIRE as of 16 Sep 2019



%\cite{Radyushkin:1983wh}
\bibitem{Radyushkin:1983wh} 
  A.~V.~Radyushkin,
  %``On Spectral Properties of Parton Correlation Functions and Multiparton Wave Functions,''
  Phys.\ Lett.\  {\bf 131B}, 179 (1983)
  %%doi10.1016/0370-2693(83)91116-4
  %%CITATION = %%doi10.1016/0370-2693(83)91116-4;%%
  %28 citations counted in INSPIRE as of 23 Jan 2018
  
  %\cite{Efremov:1978rn}
\bibitem{Efremov:1978rn} 
  A.~V.~Efremov and A.~V.~Radyushkin,
  %``Asymptotical Behavior of Pion Electromagnetic Form-Factor in QCD,''
  Theor.\ Math.\ Phys.\  {\bf 42}, 97 (1980)
  [Teor.\ Mat.\ Fiz.\  {\bf 42}, 147 (1980)]
%  doi:10.1007/BF01032111
  %%CITATION = doi:10.1007/BF01032111;%%
  %484 citations counted in INSPIRE as of 16 Sep 2019
  
  %\cite{Efremov:1979qk}
\bibitem{Efremov:1979qk} 
  A.~V.~Efremov and A.~V.~Radyushkin,
  %``Factorization and Asymptotical Behavior of Pion Form-Factor in QCD,''
  Phys.\ Lett.\  {\bf 94B}, 245 (1980)
%  doi:10.1016/0370-2693(80)90869-2
  %%CITATION = doi:10.1016/0370-2693(80)90869-2;%%
  %1097 citations counted in INSPIRE as of 16 Sep 2019
  
  %\cite{Ji:1996ek}
\bibitem{Ji:1996ek} 
  X.~D.~Ji,
  %``Gauge-Invariant Decomposition of Nucleon Spin,''
  Phys.\ Rev.\ Lett.\  {\bf 78}, 610 (1997)
%  doi:10.1103/PhysRevLett.78.610
%  [hep-ph/9603249].
  %%CITATION = doi:10.1103/PhysRevLett.78.610;%%
  %1697 citations counted in INSPIRE as of 16 Sep 2019
  
  %\cite{Mueller:1998fv}
\bibitem{Mueller:1998fv} 
  D.~M\"uller, D.~Robaschik, B.~Geyer, F.-M.~Dittes and J.~Ho{\v r}ej{\v s}i,
  %``Wave functions, evolution equations and evolution kernels from light ray operators of QCD,''
  Fortsch.\ Phys.\  {\bf 42}, 101 (1994)
 % doi:10.1002/prop.2190420202
 % [hep-ph/9812448].
  %%CITATION = doi:10.1002/prop.2190420202;%%
  %1133 citations counted in INSPIRE as of 16 Sep 2019
  
  %\cite{Radyushkin:1997ki}
\bibitem{Radyushkin:1997ki} 
  A.~V.~Radyushkin,
  %``Nonforward parton distributions,''
  Phys.\ Rev.\ D {\bf 56}, 5524 (1997)
 % doi:10.1103/PhysRevD.56.5524
%  [hep-ph/9704207].
  %%CITATION = doi:10.1103/PhysRevD.56.5524;%%
  %1073 citations counted in INSPIRE as of 16 Sep 2019
  
  %\cite{Radyushkin:1998es}
\bibitem{Radyushkin:1998es} 
  A.~V.~Radyushkin,
  %``Double distributions and evolution equations,''
  Phys.\ Rev.\ D {\bf 59}, 014030 (1999)
%  doi:10.1103/PhysRevD.59.014030
 % [hep-ph/9805342].
  %%CITATION = doi:10.1103/PhysRevD.59.014030;%%
  %258 citations counted in INSPIRE as of 16 Sep 2019
  
  %\cite{Karpie:2018zaz}
\bibitem{Karpie:2018zaz} 
  J.~Karpie, K.~Orginos and S.~Zafeiropoulos,
  %``Moments of Ioffe time parton distribution functions from non-local matrix elements,''
  JHEP {\bf 1811}, 178 (2018)
 % doi:10.1007/JHEP11(2018)178
%  [arXiv:1807.10933 [hep-lat]].
  %%CITATION = doi:10.1007/JHEP11(2018)178;%%
  %18 citations counted in INSPIRE as of 17 Sep 2019
  
  %\cite{Joo:2019jct}
\bibitem{Joo:2019jct} 
  B.~Joó, J.~Karpie, K.~Orginos, A.~Radyushkin, D.~Richards and S.~Zafeiropoulos,
  %``Parton Distribution Functions from Ioffe time pseudo-distributions,''
  arXiv:1908.09771 [hep-lat]
  %%CITATION = ARXIV:1908.09771;%%
  
  
  %\cite{Joo:2019bzr}
\bibitem{Joo:2019bzr} 
  B.~Joó, J.~Karpie, K.~Orginos, A.~V.~Radyushkin, D.~G.~Richards, R.~S.~Sufian and S.~Zafeiropoulos,
  %``Pion Valence Structure from Ioffe Time Pseudo-Distributions,''
  arXiv:1909.08517 [hep-lat].
  %%CITATION = ARXIV:1909.08517;%%
 
   \end{thebibliography}
\end{document}